\documentclass[doublecol,linenumbers]{epl2} 

\usepackage{graphicx}
\usepackage{amssymb}
\usepackage{amsmath}

\newcommand{\be}{\begin{equation}}
\newcommand{\ee}{\end{equation}}

\title{Going beyond PFA: a  precise  formula for the sphere-plate Casimir force}

\date{\today}

\author{Giuseppe Bimonte}
\institute{                    
  Dipartimento di  Fisica E. Pancini, Universit{\`a} di Napoli Federico II, Complesso Universitario
MSA, Via Cintia, I-80126 Napoli, Italy\\
   INFN Sezione di
Napoli, I-80126 Napoli, Italy
}
 

\abstract{
Quantum   fluctuations of the electromagnetic field in the medium surrounding two discharged macroscopic  polarizable bodies  induce a force between the two bodies, the so called Casimir  force. In the last two decades many experiments have accurately measured this force, and significant efforts are  made to harness it in the actuation of micro and nano machines. The inherent many body character of the Casimir force makes its computation very difficult in non-planar geometries, like the standard experimental sphere-plate configuration.  Here we derive an approximate semi-analytic formula for the sphere-plate Casimir force, which is both easy to compute numerically and very accurate at all distances. By a comparison with the fully converged exact scattering formula, we show that the error made by the approximate formula is indeed much smaller than the uncertainty of present  and foreseeable Casimir experiments.  }

\begin{document}
 
\maketitle

\begin{figure}
\includegraphics[width=.9\columnwidth]{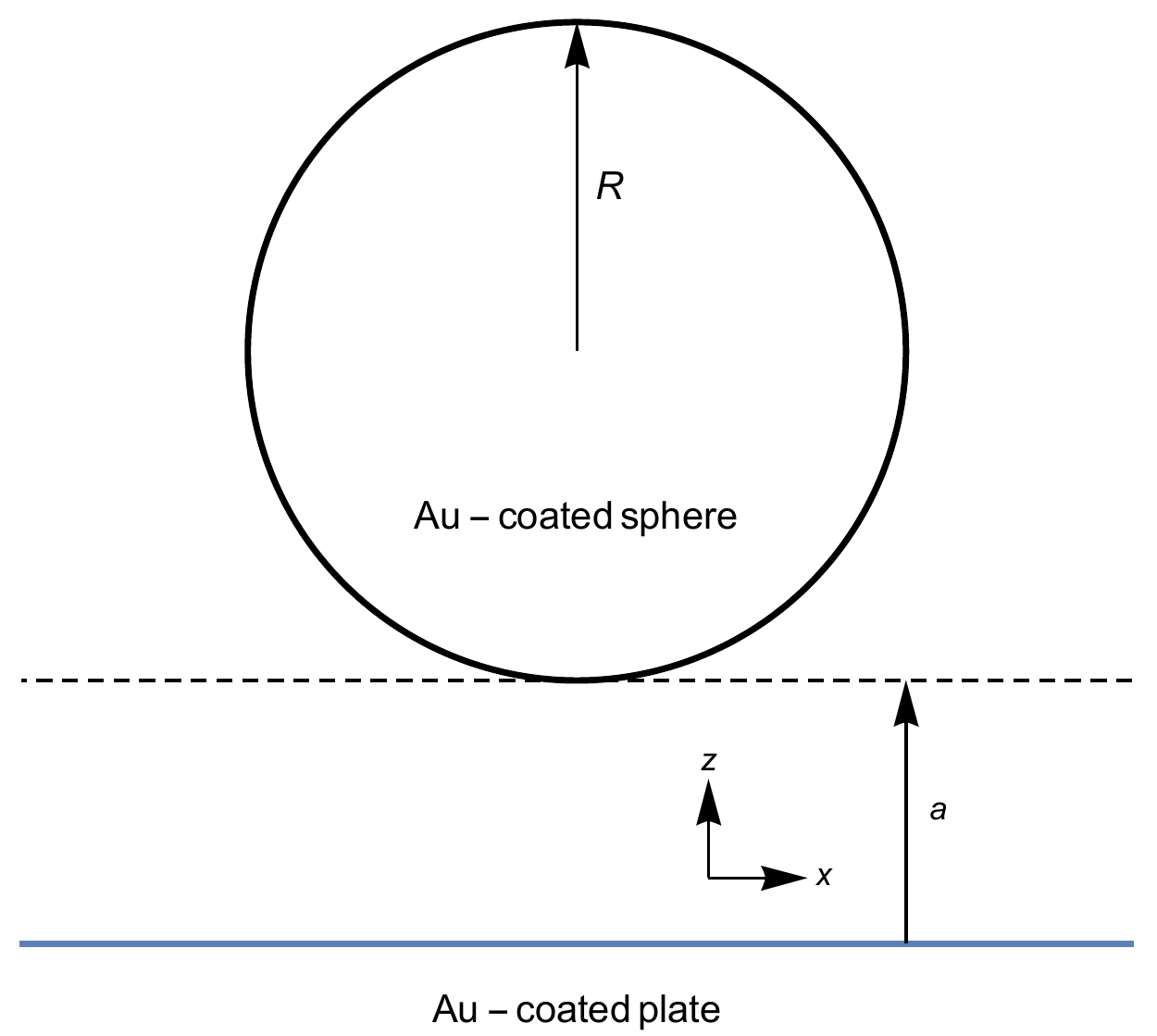}
\caption{\label{setup}  The sphere-plate Casimir setup.}
\end{figure}

The Casimir effect \cite{Casimir48} is the tiny force  acting between two (or more) discharged polarizable objects, that originates from quantum and thermal fluctuations of the electromagnetic field in the medium surrounding the bodies. It represents one of the rare manifestations of the quantum at the macroscopic scale, similar to black-body radiation, superfluidity and superconductivity. Reviews can be found in Refs.  \cite{parse,book2,milonni,lamoreaux2}. Recent years   witnessed an 
impetuous resurgence of interest in the Casimir effect, triggered by a series of precision experiments \cite{lamor1,umar} and by the exciting perspective of harnessing this force in the nanoworld \cite{capasso}. The Casimir force is notoriously difficult to compute in non-planar geometries, like the standard sphere-plate geometry  adopted in almost all experiments (see Fig. \ref{setup}). Recently, important progress in the understanding of the sphere-sphere and sphere-plate force has been made in the non-retarded or van der Waals regime by using transformation optics \cite{pendry}.  By a combination of  asymptotic techniques \cite{bordag1,bordag2,teo,fosco1,bimonte1,bimonte2} with a partial exact solution valid in the classical limit \cite{bimonteexact}, here we derive a  new semi-analytic formula for the complete retarded sphere-plate Casimir force.  Comparison with high precision numerical simulations reveals that the formula is remarkably accurate at all separations.  The new formula thus provides a  simple and yet fully reliable tool to interpret present and future experimental data. 

In  his famous 1948 paper \cite{Casimir48}, Hendrik Casimir discovered that the ground state energy of the quantized electromagnetic (em) field is modified by the presence of material  bodies that  interact with  the em field. By carefully adding up the zero-point energies of the em modes of a planar cavity consisting of two perfectly conducting  plates of (large) area $A$ at distance $a$, he obtained his celebrated formula for  the   force acting on the plates  at zero temperature:
\be
F_C=\frac{\pi^2 \hbar c}{240 \;a^4} \;A\;.
\ee 
This simple formula reveals the main features of the Casimir force: the presence of Planck's constant      
indicates its quantum character, while the presence of the speed of light $c$ shows that it is a relativistic effect. The $a^{-4}$ dependence   shows that the magnitude of the force increases rapidly as the distance $a$ is decreased: indeed, for typical experimental submicron separations the Casimir force is  much stronger than gravity. 

An important extension of Casimir's work was made by E.~Lifshitz in 1955 \cite{lifs}:  using  Rytov's theory of electromagnetic fluctuations \cite{rytov}, Lifshitz  worked out the Casimir pressure between two real dielectric parallel plates described by the respective (complex) dynamical permittivities $\epsilon(\omega)$, at a finite temperature $T$. Lifshitz' results paved the way  to investigations of the Casimir effect in real physical conditions.   

Even though the first observation of the Casimir force was reported just a few years after Casimir's prediction \cite{sparnaay}, the modern era of Casimir physics began only in the late 1990's \cite{lamor1,umar}, when a series of  experiments obtained the first  precise measurements of the force.

Despite the plane-parallel geometry studied by Casimir and Lifshitz enormously simplifies  the theoretical  analysis,  plane-parallel plates are never used in experiments (with very few notable exceptions \cite{bressi}) because it is very hard to precisely align two planar surfaces placed at a submicron distance from each other. To avoid this difficulty,  almost all experimental groups  \cite{lamor1,umar,decca6,liq,iannuzzi,palasantzas,deccaNi} adopt the sphere-plate geometry (see Fig. \ref{setup}), which obviously does not suffer from the parallelism issue. In order to  increase the strength of the Casimir force,  as it is necessary for a precise measurement to be possible,    experiments use  setups with spheres of radii $R> 10 \;\mu$m with  large aspect ratios  $R/a$  typically exceeding $10^2 \div 10^3$.

Unfortunately, the sphere-plate (and more in general  any non-planar) geometry presents the serious drawback that the computation of the Casimir force gets very difficult.  Indeed, the Casimir force is non-additive due to its inherent many-body character and therefore  its dependence  on the system geometry  is very complicated. This explains why still today the  universal tool  used to interpret  the experiments is the old-fashioned Derjaguin \cite{Derjaguin}  Proximity Force Approximation (PFA), which expresses the Casimir force between two gently curved surfaces as the average of the  force for two parallel plates,  taken over the local surface-surface separation.

This state of things has not changed despite the theoretical breakthrough in the early 2000's \cite{sca1,kenneth,sca2}, when  using  scattering methods    a {\it mathematically exact}  formula for the Casimir interaction between  two (or more) compact bodies of any shape has been derived, in the form of a multipole expansion in terms  of the respective T-matrices. In principle, the   scattering formula  allows   to exactly compute  the Casimir force between two objects whose T-matrices are either known, or can be worked out numerically.  The experimental sphere-plate geometry is among these, since  its T-matrix  has been known for a very long time \cite{jackson}.  Notwithstanding this, in the past the   scattering formula has never been used in practice, because its slow convergence  makes it  difficult  to  compute  it  in a reasonable computer time for the large aspect ratios $R/a$ of the experiments (more on this below). 
 Only very recently   a large-scale simulation of the sphere-plate scattering formula going up to  multipole order $l_{\rm max}=2 \times 10^4$  appeared in the arXiv \cite{Ingold}, in which  the sphere-plate force and force gradient have been computed numerically for experimentally  relevant aspect ratios $R/a \sim 4 \times 10^3$.

The  PFA has been  put on a firmer basis recently, by showing   rigorously that it coincides with the leading term of  the asymptotic  expansion of the exact scattering formula for large $R/a$ (at fixed $a$) \cite{bordag1}. By the same approach, it has been also possible to estimate the $O(a/R)$ corrections to the PFA \cite{bordag2,teo}. An alternative and equivalent method to  compute  curvature corrections to the PFA    is based on  a derivative expansion (DE) of the Casimir energy  in powers of derivatives of the curved surface height profile \cite{fosco1,bimonte1,bimonte2}.

Before we plunge into computations, it is useful to discuss shortly   why computing precisely the Casimir force is  important. There are at least two reasons for that. The first reason has to do with Casimir physics, while the second  is connected with recent searches on non-newtonian gravity in the sub-millimiter range. Let us consider the former first. 

The most precise recent   experiments   utilizing Au-coated sphere and plate,   have measured the Casimir force and its gradient with experimental errors around one percent, for separations smaller than a few hundred nm.  The theoretical analysis of these experiments revealed perplexing discrepancies, at the percent level, with predictions based on the PFA. The interpretation of these discrepancies spurred a heated debate in the community. It has been argued by some authors that these discrepancies point  at a fundamental flaw in Lifshitz theory, regarding the role of free charge carriers in determining the magnitude of the thermal component of the Casimir force \cite{book2}. This debate has become known in the Casimir field as the Drude vs. plasma controversy \cite{brevik}: in essence the controversy is whether in Lifshitz formula one should use the familiar Drude model on rather the dissipationless plasma model of infra-red optics  to describe the response of  a metallic plate at low frequencies. On physical grounds one would expect that the plasma model should be used to compute the Casimir interaction between two superconducting plates \cite{bimontesuper}, while for normal metals at room temperature as those used in the experiments the Drude model should provide the correct description. Contrary to general expectations,  a series of experiments shows that  the plasma model is in agreement with data, while the physically plausible Drude model is not \cite{decca6,17,18,19,21,22,23}. There is however a single experiment which is in agreement with the Drude model \cite{25}. On the theory side, it has been realized that use of  the Drude prescription in Lifshitz formula also leads to violation of Nernst heat theorem, in the idealized limit of metal test bodies with perfect lattice structures \cite{10,11,12}.    Since the problem  has a bearing on the fundamental principles of statistical physics, many feel desirable to make sure that the PFA, which has been used to interpret the experiments,  is  precise enough. 

The ability to compute precisely the Casimir force is also  crucial in recent searches for non-newtonian gravity in the sub-millimiter range \cite{decca6,deccayuk}.  Here, one measures the force of attraction between two test bodies, looking for deviations from Newton's law.  Usually  the test bodies used in these experiments  are Au-coated, to avoid potentially harmful forces due to stray electrostatic fields \cite{25}. When separations in the micron and sub-micron region are probed,  the measured force is undistinguishable within the errors from the  Casimir force, which is much stronger than the Newtonian force, and so one can only set bounds on hypotetical non-newtonian forces of the Yukawa type \cite{decca6,deccayuk}. The strength of the obtained bounds depends crucially on the precision of the theoretical prediction of the Casimir force.

The above considerations motivated us to see if it is possible to work out a formula for the sphere-plate  Casimir force that combines the precision of the scattering formula with the computational simplicity of the PFA. Our work builds up on the important  progress that has been made in recent years in understanding the properties of the scattering formula.  

We consider a sphere of radius $R$ placed at minimum distance $a$ from a plane (see Fig. \ref{setup}). For simplicity we  assume that the sphere and plate are made of the same material, characterized by a common dynamical complex permittivity $\epsilon(\omega)$. We shall specifically consider the case of Au, which is the standard material used in modern Casimir experiments.  The exact scattering formula \cite{sca1,kenneth,sca2} for the Casimir  free energy reads
\be
{\cal F}=k_B T \sum_{n \ge 0}\;\!\!' \;{\rm Tr} \ln[1-\hat{M}(\rm {i} \xi_n)]\;,\label{freeen}
\ee  
where $k_B$ is Boltzmann constant, $T$ is the temperature, $\xi_n= 2 \pi n k_B T/\hbar$  are the (imaginary) Matsubara frequencies, the prime sign in the sum indicates that the $n=0$ term is taken with weight 1/2.  The trace ${\rm Tr}$ in this equation is taken over the spherical multipoles indices $(l,m)$ and on the  polarization indices $\alpha={\rm TE,TM}$:
\be
{\rm Tr}= \sum_{m=-\infty}^{\infty} \sum_{l={\rm max} \{1,|m|\}}^{\infty} {\rm tr}\;,
\ee  
where ${\rm tr}$ denotes the trace over $\alpha$. The matrix elements  $M_{lm\alpha,m'l'\alpha'}$ of ${\hat M}$ shall not be reported here for brevity. Their explicit expressions can be found for example in Ref.\cite{teo}.  It suffices to say here that  $M_{lm\alpha,m'l'\alpha'}$ involves the Fresnel reflection coefficients of the plane, and the Mie scattering coefficients of the sphere, both evaluated for the imaginary Matsubara frequencies ${\rm i} \xi_n$. The Fresnel coefficients are  multiplied by suitable products of the associated Legendre polynomial $P^m_l(\cosh(\theta))$ and $P^{m'}_{l'}(\sinh(\theta))$, or their derivatives, and then integrated over the (imaginary) incidence angle $\theta$.  The expressions for the Casimir force $F=-d {\cal F}/da$ and its gradient $F'= d F /d a$  are obtained by taking  derivatives of Eq. (\ref{freeen}) with respect to the separation $a$. In numerical simulations, the multipole indices $l,l',m,m'$ and the Matsubara index $n$ are truncated to $l,l' \le l_{\rm max}$, $m\le m_{\rm max}$ and $n \le n_{\rm max}$. The values of $l_{\rm max}, m_{\rm max}$ and $n_{\rm max}$  for which convergence is achieved scale with the separation $a$ as $l_{\rm max} \sim R/a$, $m_{\rm max} \sim \sqrt{R/a}$   and $n_{\rm max} =\lambda_T/ a$, where $\lambda_T=\hbar c/2 \pi k_B T=1.2 \;\mu$m is the thermal length \cite{bordag1,bordag2,teo}. It turns out that to estimate the Casimir force and its gradient with an accurary of $10^{-4}$, as we seek,  one needs $l_{\rm max} \simeq 6 R/a$, $m_{\rm max} \simeq 6 \sqrt{R/a}$, $n_{\rm max} \simeq 10\, \lambda_T/a$.  Previous simulations \cite{Neto,Neto2} with $l_{\rm max}= 45$ were used to compute quite precisely the force for $R/a < 20$. As we mentioned earlier, a new large simulation \cite{Ingold} reached $l_{\rm max}= 2 \times 10^4$,  making it possible to probe aspect ratios up to $R/a \sim 4 \times 10^3$. In our simulations we went up to $l_{\rm max}= 120$, which allows us to compute  the force with a precision of $10^{-4}$ for $R/a \le  20$.
 
Below we prove that a  very accurate formula for the Casimir force and its gradient   can be obtained by a combined use of certain limiting exact solutions \cite{bimonteexact}  and asymptotic formulae \cite{bordag1,bordag2,teo,bimonte2} that have been  recently reported in the literature. 

In \cite{bimonteexact}  the separability of the Laplace Equation in bispherical coordinates was exploited to derive an {\it exact} analytical formula for the classical limit,  i.e. the $n=0$ term of the complete scattering formula Eq. (\ref{freeen}),  of the Casimir energy of two metallic spheres described by the Drude model. Unfortunately, no exact solution is yet available for the $n=0$ mode for the plasma model. Bispherical coordinates have been also used in \cite{pendry} to derive a fast convergent numerical scheme  and an approximate analytical formula for the van der Waals interaction of metallic plasmonic spheres.  Another application of bispherical coordinates can be found in the work \cite{bimonteN}, which deals with the classical Casimir interaction of perfectly conducting sphere and plate. In the special sphere-plate case, the exact  formula derived in \cite{bimonteexact} for the Drude classical Casimir energy reads: 
$$
{\cal F}_{n=0}^{(\rm exact)}=\frac{k_B T}{2} \left\{\sum_{l=1}^{\infty} (2 l +1) \ln[1-Z^{2 l+1}]\;\right.
$$
\be
\left.+\ln \left[1-(1-Z^2) \sum_{l=1}^{\infty} Z^{2 l+1} \frac{1-Z^{2 l}}{1-Z^{2 l+1}} \right] \right\}\;,\label{enerDr}
\ee
where the parameter $Z$ depends on the inverse of the aspect ratio  $x=a/R$:
\be
Z=[1+x+\sqrt{x\,(2+x)}]^{-1}\;.\label{Zdef}
\ee
By taking   a derivative of the above formula with respect to $a$,   the corresponding expression for the Casimir force $F_{n=0}^{(\rm exact)}$ and its derivative ${F'}_{n=0}^{(\rm exact)}$ are easily obtained. 

Now we turn to the $n>0$ terms of the scattering formula Eq. (\ref{freeen}). Unfortunately, for  $n>0$  no exact solution is yet available, and we have to make recourse to approximations. It turns out that the contribution ${ F}_{n>0}$ of the $n>0$ terms to the Casimir  force can be computed very accurately  using the recently proposed  DE \cite{fosco1,bimonte1,bimonte2}.  To introduce the DE, we consider instead of the sphere a more general gently curved dielectric surface, described by a smooth height profile $z=H(x,y)$, where $(x,y)$ are cartesian coordinates spanning the plate surface $\Sigma$, and the $z$ axis is drawn perpendicular to the plate towards the surface.   The starting point of the  DE is the assumption that the functional ${F}[H]_{n>0}$ admits a  {\it local} expansion in {\it powers of derivatives} of the  height profile $H$: 
\be
{ F}[H]_{n>0}={ F}_{n>0}^{(\rm PFA)}[H] + \int_{\Sigma}  d^2 x \; \alpha(H)  (\nabla H)^2 + \rho^{(2)}\;.\label{derexp}
\ee The first term on the r.h.s. of the above equation coincides with the  PFA (restricted to modes with $n>0$)
\be
{ F}_{n>0}^{(\rm PFA)}[H]= \sum_{n>0}\int_{\Sigma}  d^2 x \;F^{(\rm pp)}_n(H)\;,
\ee
where $F_n^{(\rm pp)}(H)$ represents the contribution of the $n$-th Matsubara mode to the unit-area Casimir force between two parallel plates at distance $H$, as given by Lifshitz formula \cite{lifs}. The coefficient
  $\alpha(H)$ is a function to be determined, while the quantity  $\rho^{(2)}$ represents corrections that becomes negligible as the local radius of curvature of the surface $R$ goes to infinity  for fixed minimum surface-plate distance $a$. The validity of the ansatz made in Eq. (\ref{derexp}) depends essentially on the locality properies of the Casimir force. The key point to notice here is that for imaginary frequencies $\omega={\rm i} \xi_n$ with $n>0$  the photons aquire an effective mass proportional to $n$, which renders the interaction more and more local as $n$ increases,  thus making the DE Eq. (\ref{derexp}) more and more accurate.  In previous works \cite{bimonte1,bimonte2} the DE was used for {\it all modes}, including $n=0$. For $n=0$ the exact sphere plate energy includes at the sub-sub-leading order complicated logarithmic terms \cite{bimonteexact} which render the DE less accurate for realistic values of $R/a$. 
Restriction of the DE to the massive $n>0$ modes is the key ingredient  to achieve a high precision.  
The function $\alpha(H)$  is determined by matching the DE with the perturbative expansion \cite{lambrechtpert} of the Casimir force, in the common domain of validity of the two expansions (for details, see Refs. \cite{fosco1,bimonte1,bimonte2}). When the integral in Eq. (\ref{derexp}) is evaluated for a sphere with $H(x,y)=a+(x^2+y^2)/(2 R)+ \dots$,  and only terms up to  order $a/R$ are retained, one ends up with an expression that can be recast in the form:
\be
{ F}_{n>0}={ F}_{n>0}^{(\rm PFA)} \left(1- \theta \; \frac{a}{R} \right)\;.\label{DEF}
\ee
In this equation ${ F}_{n>0}^{(\rm PFA)}$ coincides with the familiar PFA formula for the sphere-plate force (restricted to modes with $n>0$) 
\be
{ F}_{n>0}^{(\rm PFA)}= 2 \pi R \sum_{n>0} {\cal F}_{n}^{(\rm pp)}\;,
\ee
where $ {\cal F}_{n}^{(\rm pp)}$ denotes the contribution of the $n$-th Matsubara mode to the unit-area free-energy of two parallel plates, as given by Lifshitz formula \cite{lifs}.
An analogous expression is obtained for the force gradient $F'$:
\be
{ F'}_{n>0}={ F '}_{n>0}^{(\rm PFA)}\left(1 -\tilde{ \theta} \; \frac{a}{R} \right)\;,\label{DEFpr}
\ee
where ${ F '}_{n>0}^{(\rm PFA)}=d { F}_{n>0}^{(\rm PFA)}/d a $. 
The coefficients $\theta$ and ${\tilde \theta}$ both  depend on $a, \lambda_T$ and on the length parameters characterizing the materials of the sphere and plate (in the case considered here the plasma length $\lambda_p$ of Au), but are independent of the sphere radius $R$.  In table I we list   $\theta$  and $\tilde{\theta}$ calculated for Au at room temperature ($T= 300$ K),  using tabulated optical data \cite{palik}. The weighted Kramers-Kronig dispersion relations \cite{bimonteKK} was used to compute precisely $\epsilon({\rm i} \xi_n)$ starting from the real-frequency optical data given by Palik. The coefficients $\theta$  and $\tilde{\theta}$ can be calculated also using the method developed by Bordag and Nikolaev \cite{bordag1,bordag2,teo}, and we have verified that this alternative method  gives for $\theta$ and ${\tilde \theta}$ the same results as the DE. 

Combining Eq. (\ref{enerDr}), with Eq. (\ref{DEF})  we end up with the following approximate expression for $F$:
\be
{ F}_{\rm approx}={ F}_{n=0}^{(\rm exact)}+{ F}_{n>0}^{(\rm PFA)} \left(1- \theta \; \frac{a}{R} \right)\;,\label{for}
\ee
An analogous formula can be derived for the force gradient $F'$:
\be
F'_{\!\rm approx}={ F'}_{n=0}^{(\rm exact)}+{ F'}_{n>0}^{(\rm PFA)} \left(1- \tilde{\theta} \; \frac{a}{R} \right)\;.\label{derfor}
\ee
The above two Equations represent the main result of this work. 
\begin{largetable}
\caption{Values of the coefficients $\theta$ and ${\tilde \theta}$ for Au at room temperature.}
\label{tab.1}
\begin{center}
\begin{tabular}{ccccccccccccc} \hline
$a (\mu m)$\;\; &0.10& 0.15 & 0.2  \;\;& 0.25\;\;\;& 0.3\;\; &0.35 \;\;& 0.4 \;\; & 0.45 & 0.5\;\;& 0.55 & 0.6\;\; & 0.65  \\ \hline \hline
$\theta$ \;\;&0.717 &0.694 &\; 0.6645\;\; & 0.636 & 0.609 \;\;&0.584 & 0.561\;\;&0.540  & 0.520\;\;&0.502 & 0.485\;\; &0.468 \;\;    \\  \hline
${\tilde \theta}\;\;$& 0.456  &0.4715 & 0.470\;\; &0.463 & 0.454 \;\;&0.4445 & 0.435 \;&0.425 & 0.415 &0.4055 & 0.396& 0.387    \\ \hline \\
 \hline
$a (\mu m)$\; &0.70&0.75 & 0.8 & 0.85 & 0.9& 0.95&  1 &1.2&1.4 &1.6 & 1.8 & 2  \\ \hline \hline
$\theta$ \;\; &0.453&\;0.439 \;&0.425 &0.413&0.400&0.389& 0.378\;\;&0.339&0.307&0.279&0.256&0.237 \\  \hline
${\tilde \theta}\;\;$&0.379 &0.370  &0.362 &0.3545 &0.347&0.3395&0.332 &0.306&0.282&0.261& 0.242 &0.225   \\ \hline  \hline
\end{tabular}
\end{center}
\end{largetable}

\begin{figure}
\includegraphics[width=.9\columnwidth]{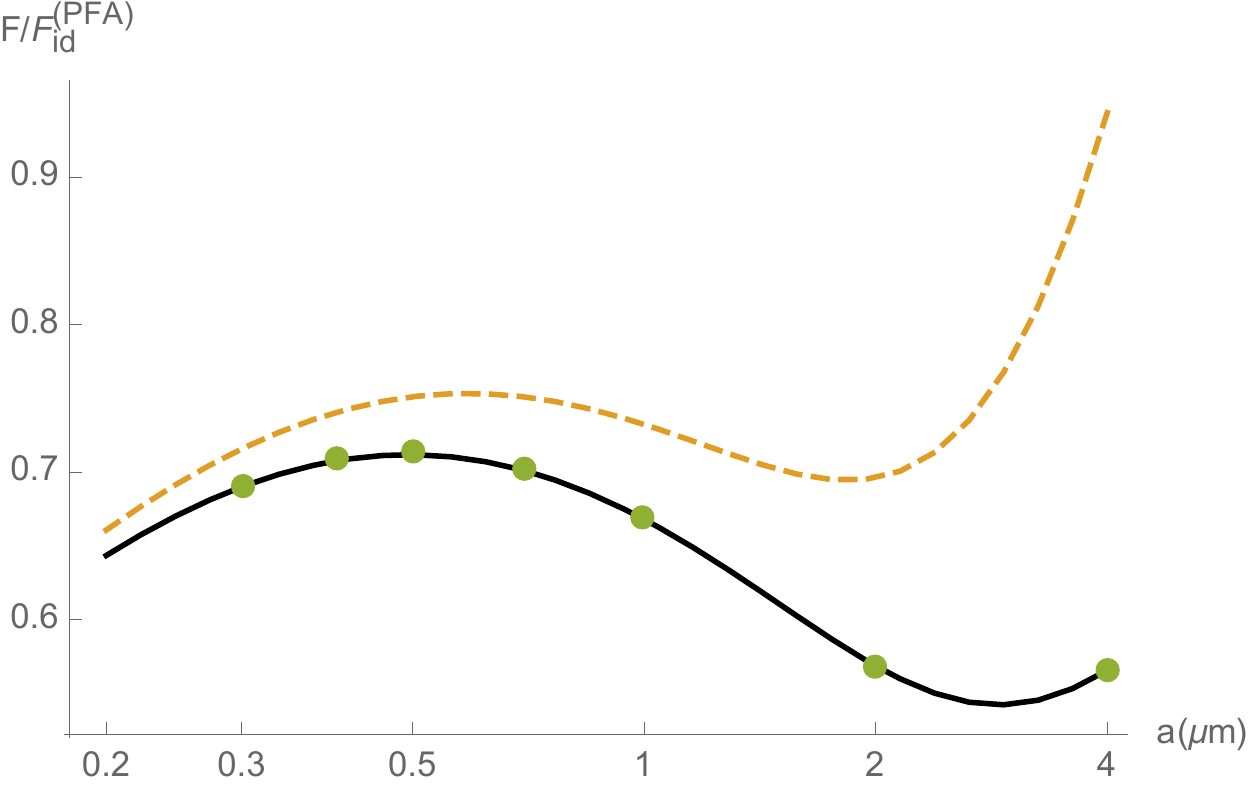}
\caption{\label{plot1}  Casimir force for Au  sphere ($R=5\;\mu$m) and plate at room temperature, normalized by the PFA force for ideal plates. The dots are numerical data computed using the fully converged exact scattering formula, the solid line is for the approximate formula Eq. (\ref{DEF}), while the dashed line shows the standard PFA. All forces were computed for room temperature ($T=$ 300 K) using Palik's optical data for Au \cite{palik}, extrapolated to low frequencies by the Drude model.}
\end{figure}

\begin{figure}
\includegraphics[width=.9\columnwidth]{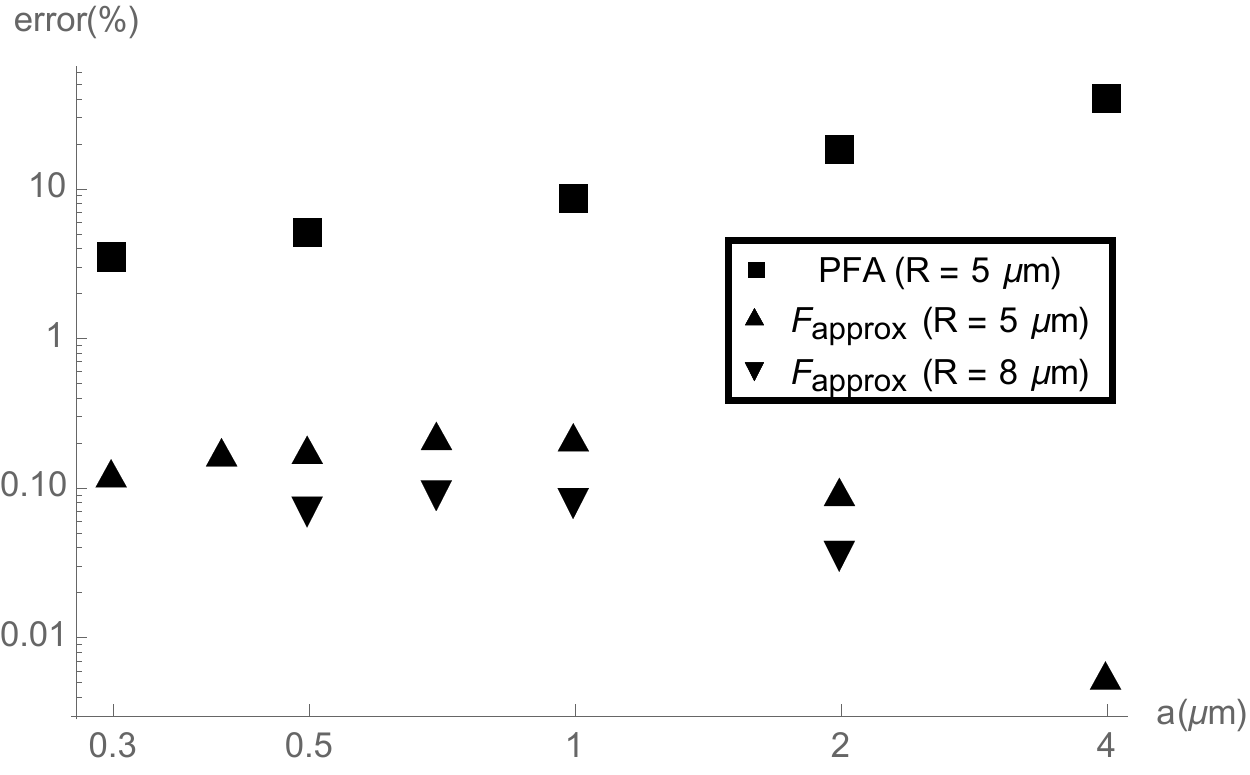}
\caption{\label{errors}   Percent errors on the Casimir force made by the PFA (squares) and by the formula in Eq. (\ref{for}) (triangles). }
\end{figure}
We tested the precision of Eqs. (\ref{for}) and (\ref{derfor}) by comparing them with high precision numerical simulations up to $l_{\rm max}=120$ of the exact  force and force gradient derived from the scattering formula Eq.(\ref{freeen}).  In our simulations the scattering formula was used only to estimate the modes $n>0$, while for $n=0$ we used the exact analytical formula Eq. (\ref{enerDr}). In Fig. \ref{plot1} we show the Casimir force for a Au-sphere of radius $R=5\;\mu$m. The force has been normalized by the PFA force for ideal sphere and plate $F_{\rm id}^{\rm (PFA)}=- \pi^3 \hbar c R/(360 a^3)$. The dots represent the numerical data computed using the fully converged exact scattering formula. The solid line shows a plot of our approximate formula for the force Eq. (\ref{DEF}), while the dashed line is  the standard PFA. It is evident that the PFA  performs fairly well for small separations, as expected, while it does quite poorly for moderate values of $R/a$.  On the contrary   Eq. (\ref{DEF}) works remarkably well at all separations.  A clear demonstration of the precision of Eq. (\ref{DEF})  is provided by Fig. \ref{errors}  which compares the percent errors made by the PFA (squares)  and by  Eq. (\ref{DEF}) (triangles). For $R= 5\;\mu$m  the maximum error made by Eq. (\ref{DEF})  is of 0.2 percent for $a=1\;\mu$m. The fact that the error becomes very small for separations $a$ exceeding a few microns should not come as a surprise, because  for separations $a \gg\lambda_T=1.2\; \mu$m the exact classical $n=0$ term dominates. More remarkable is the excellent performance of the formula on the small-distance side.   The nature of the DE makes one expect that our formula becomes more and more accurate for smaller separations and/or larger sphere radii. Both expectations are fully confirmed by the data:   when $a$ is decreased from one micron to 300 nm, keeping $R= 5\;\mu$m, the error decreases from 0.2 to to 0.1 percent. At the same time, when $R$ is increased from five to eight microns and $a$ is kept fixed, the error is seen to decrease for all separations. For example, for $a= 1\,\mu$m, the maximum error decreases from 0.2 percent to 0.08 percent.   

The dynamical experiment in \cite{decca6} used a microtorsional mechanical oscillator to measure  precisely the Casimir force derivative $F'$ between a Au sphere and plate for separations smaller than 700 nm, with an error of one percent for the smallest probed separation of 162 nm. The theoretical analysis in that experiment was based on the PFA, and it was concluded that the data are in disagreement with the Drude prescription, while they are in agreement with the   plasma prescription \cite{book2}.  It is interesting to compare the Drude-prescription values of the force-derivative $F'$  provided by our Eq. (\ref{derfor}) with the PFA. 
For the aspect ratios considered here, a conservative estimate of the error made by $F'_{\rm approx}$ is smaller than 0.08 percent. In Table II we list the respective values of $F'/ 2 \pi R $ in mPa, computed for $R= 150\;\mu$m. Within the PFA,  $F_{\rm PFA}'/ 2 \pi R = F_{pp}$ where $F_{pp}$ is the plane-parallel pressure as given by Lifshitz formula \cite{lifs}.
 As we see,  $F'_{\rm approx}$ differs from the PFA by less than 0.2 percent,  confirming that the PFA is   adequate for this experiment. The recent numerical simulation in \cite{Ingold} shows that   the error made by the PFA is  about two times larger if the plasma prescription is used, instead of the Drude prescription considered here.
\begin{largetable}
\begin{tabular}{ l ll l l l l} \hline
$a (\mu m)$\;\; & 0.2  \;\;& 0.3\;\;  & 0.4 \;\; & 0.5\;\;& 0.6 & 0.7 \\ \hline \hline
${F'}_{\!\!\rm approx}/2 \pi R$ \;\;\;& 493.4  & 109.0  & 36.50  & 15.415  & 7.557 &4.110  \\ \hline
${F'}_{\!\!\rm PFA}/2 \pi R$\;\;\;\;  & 493.7  & 109.1  & 36.53  & 15.43  & 7.568 & 4.117 \\  \hline
\end{tabular}
\caption{ Values of  $F'/2 \pi R$ (in mPa) for a Au sphere-plate at room temperature, computed using Eq. (\ref{derfor}) and the PFA. }
\end{largetable} 

In this letter we have derived a semi-analytic formula for the Casimir force and its gradient in the standard sphere-plate geometry used in  current precision experiments.   Although approximate, Eqs. (\ref{DEF}) and (\ref{DEFpr}) are very accurate at all separations.  Comparison  with high precision numerical simulations of the exact scattering formula demonstrates that the precision of the approximate formulae far exceeds the experimental uncertainty of present and foreseeble experiments, to the extent that  the approximate formulae can be  considered as experimentally indistinguishable from the exact scattering formula. In this work we restricted our attention to the Drude prescription. Extension  to the plasma model requires non-trivial modifications only for the  $n=0$ mode, since in the plasma case no exact solution is yet available for this mode.  For the $n>0$ modes Eqs. (\ref{DEF}-\ref{DEFpr}) remain valid, apart from a minor change in the numerical values of the coefficients $\theta$ and $\tilde \theta$.  We leave this topic   for a future work.

It would be of great interest to carry out a detailed comparison between the predictions obtained by the  large-scale simulation  of \cite{Ingold} and the formulae presented here.

\acknowledgments

The author thanks T. Emig, N. Graham, R. Jaffe and M. Kardar for discussions.


\begin{thebibliography}{200}

\bibitem{Casimir48}  Casimir H.~B.~G., {\it Proc. K. Ned. Akad. Wet.}, {\bf 51} (1948) 793.

\bibitem{parse}  Parsegian V. A., {\it Van der Waals Forces} (Cambridge University Press, Cambridge, England,
2005).


\bibitem{book2}  Bordag M.,  Klimchitskaya G.L.,  Mohideen U. and  Mostepanenko V.M.,  {\it Advances in the Casimir Effect} (Oxford University Press, Oxford, 2009).

\bibitem{milonni} P. Milonni, {\it The Quantum Vacuum: An Introduction to Quantum Electrodynamics} (Academic, 1994).



\bibitem{lamoreaux2}  Lamoreaux S.K., {\it Phys. Today}, {\bf 60} (2007) 40.

\bibitem{lamor1}  Lamoreaux S.K., {\it Phys. Rev. Lett.}, {\bf 78} (1997) 5.

\bibitem{umar}   Mohideen U. and  Roy A., {\it Phys. Rev. Lett.}, {\bf 81} (1998) 4549.

\bibitem{capasso}  Chan H.B.,  Aksyuk V.A.,  Kleiman R.N., Bishop D.J.  and  Capasso F., {\it Science}, {\bf 291}, (2001) 1941. 

\bibitem{pendry} Pendry  J. B.,  Fern\'andez-Dom\'inguez A. I., Yu Luo and  Zhao R., {\it Nat. Phys.}, {\bf 9} (2013) 518.

\bibitem{bordag1} Bordag  M.  and  Nikolaev V.,  {\it J. Phys. A}, {\bf 41} (2008)164002.

\bibitem{bordag2} Teo L. P.,  Bordag M. and Nikolaev V. , {\it Phys. Rev. D}, {\bf 84} (2011)125037.
 
\bibitem{teo}   Teo L. P., Phys. Rev. D {\bf 88}, 045019 (2013).

\bibitem{fosco1} Fosco C. D., Lombardo  F. C. and  Mazzitelli F. D.,  {\it Phys. Rev. D}, {\bf 84} (2011) 105031.

\bibitem{bimonte1}  Bimonte G., Emig T. , Jaffe R.L.  and  Kardar M.,  {\it EPL}, {\bf 97} (2012) 50001.

\bibitem{bimonte2}  Bimonte G.,  Emig T.   and M. Kardar,  {\it Appl. Phys. Lett.} {\bf 100}, (2012) 074110.

\bibitem{bimonteexact}  Bimonte G., Emig  T.,   {\it Phys. Rev. Lett.},
{\bf 109} (2012) 160403.

 
\bibitem{lifs}  Lifshitz E. M., {\it Zh. Eksp. Teor. Fiz.}, {\bf 29} (1955) 94 [Sov. Phys. JETP {\bf 2}, 73 (1956)].

\bibitem{rytov}  Rytov S.M., {\it Theory of Electrical Fluctuations and Thermal Radiation} (Publyshing House, Academy os Sciences, USSR,1953).

\bibitem{sparnaay}  Sparnaay M., {\it Physica}, {\bf 24} (1958) 751.

\bibitem{bressi}  Bressi G.,  Carugno G., Onofrio R.  and Ruoso  G., {\it Phys. Rev. Lett.}, {\bf 88} (2002) 041804.

\bibitem{decca6} Decca  R. S.,   L\'opez D.,  Fischbach E. et al, {\it Eur. Phys. J. C}, {\bf 51} (2007)  963.

\bibitem{liq}  Munday J. N.,  Capasso F.  and  Parsegian V. A.,  {\it Nature}, {\bf 457} (2009) 170.

\bibitem{iannuzzi}  de Man S.,  Heeck K., Wijngaarden R. J.  and  Iannuzzi D., {\it Phys. Rev. Lett.}, {\bf 103} (2009) 040402.

\bibitem{palasantzas} Sedighi M., Svetovoy  V. B. and  Palasantzas G., {\it Phys. Rev. B}, {\bf 93} (2016) 085434.

\bibitem{deccaNi}  Bimonte G.,  L\'opez D. and  Decca R.S., {\it Phys. Rev. B}, {\bf 93} (2016) 184434. 

\bibitem{Derjaguin}Derjaguin  B. , {\it Kolloid Z.}, {\bf 69} (1934) 155.

\bibitem{sca1} Lambrecht A., Maia Neto P. A.  and  Reynaud S.,  {\it New J. Phys.},
{\bf 8} (2006) 243.

\bibitem{kenneth} Kenneth O.  and  Klich I., {\it Phys. Rev. Lett.} {\bf 97} (2006)160401; {\it Phys. Rev. B},  {\bf 78} (2008) 014103.

\bibitem{sca2} Emig T.,  Graham N.,  Jaffe R. L. and Kardar M. , {\it Phys. Rev. Lett.},
{\bf 99} (2007) 170403.

\bibitem{jackson} J. D. Jackson {\it Classical Electrodynamics} (John Wiley \& Sons, New York 1999).

\bibitem{Ingold} Hartmann M., Ingold G.-L. and Maia Neto P.~A., arXiv:1705.04196v1.

\bibitem{brevik} Brevik I. ,  Ellingsen S. A. and Milton K. A., {\it New J. Phys.}, {\bf 8} (2006) 236.

\bibitem{bimontesuper} G. Bimonte, {\it Phys. Rev. A} {\bf 78} (2008) 062101.

\bibitem{17} Decca R.~S.~,  Fischbach E.~, Klimchitskaya G.~L.~, Krause D.~E.,
L\'opez D. and Mostepanenko V.~M., {\it Phys. Rev. D}, {\bf 68} (2003) 116003.

\bibitem{18}
Decca R.{\ }S., L\'opez  D.,  Fischbach E.,  Klimchitskaya G.{\ }L.,
E. Krause D. and  Mostepanenko V.M., {\it Ann. Phys. (N.Y.)}, {\bf 318} (2005)
37.

\bibitem{19}
Decca R.~S., L\'opez  D.,  Fischbach E.,  Klimchitskaya G.L.,
Krause D.~E. and Mostepanenko V.~M.,
{\it Phys. Rev. D}, {\bf 75} (2007) 077101.

\bibitem{21}
Chang C.-C.,Banishev  A.~A., Castillo-Garza R.,
 Klimchitskaya G.L.,  Mostepanenko V.M. and Mohideen U.,
{\it Phys. Rev. B}, {\bf 85} (2012) 165443.

\bibitem{22}
Banishev A.~A.,
Klimchitskaya G.~L.,Mostepanenko V. M. and Mohideen U.,
{\it Phys. Rev. Lett.},  {\bf 110} (2013) 137401.

\bibitem{23}
Banishev A.~A.,
Klimchitskaya G.~L.,  Mostepanenko V. M. and Mohideen U.,
{\it Phys. Rev. B},  {\bf 88} (2013)155410.

\bibitem{25}
 Sushkov A.~O., Kim W.~J.,  Dalvit D. A. R.
and Lamoreaux  S. K.,
{\it Nature Phys.},  {\bf 7} (2011) 230.



\bibitem{10}
Bezerra V. B,  Klimchitskaya G.L. and  Mostepanenko V. M.,
{\it Phys. Rev. A},  {\bf 65} (2002) 052113.

\bibitem{11}
Bezerra V.{\ }B., Klimchitskaya  G.{\ }L. and Mostepanenko V.{\ }M.,
{\it Phys. Rev. A}, {\bf 66} (2002) 062112.

\bibitem{12}
V.{\ }B.\ Bezerra, G.{\ }L.\ Klimchitskaya, V.{\ }M.\ Mostepanenko,
and C.\ Romero,
{\it Phys. Rev. A}, {\bf 69} (2004) 022119.

\bibitem{deccayuk} Decca R. S. ,  L\'opez D., Fischbach E. , Klimchitskaya G. L. ,  Krause D. E. and  Mostepanenko V. M., {\it Phys. Rev. D}, {\bf 74} (2007) 077101.

\bibitem{bimonteN} G. Bimonte, {\it Phys. Rev. D}, {\bf 95} (2017) 065004.

\bibitem{Neto}  Canaguier-Durand A., Maia Neto  P. A., Cavero-Pelaez I. ,  Lambrecht A. and  Reynaud S.,
{\it Phys. Rev. Lett.}, {\bf 102} (2009) 230404.

\bibitem{Neto2}  Canaguier-Durand A., Maia Neto  P. A. , Lambrecht A.  and Reynaud  S.,
{\it Phys. Rev. A}, {\bf 82} (2010) 012511.

\bibitem{lambrechtpert}  Maia Neto P. A.,  Lambrecht A. and  Reynaud S., {\it Phys. Rev. A}, {\bf 72} (2005) 012115.

\bibitem{palik}  {\it Handbook of Optical Constants of Solids},
edited by E. D. Palik (Academic, New York, 1995).

\bibitem{bimonteKK} Bimonte G. , {\it Phys. Rev. A}, {\bf 83} (2011) 042109.



\end{thebibliography}
\end{document}